# Curved detectors for wide field imaging systems: impact on tolerance analysis

Eduard Muslimov[1]*[a,b], Emmanuel Hugot[a], Simona Lombardo[a], Melanie Roulet[a], Thibault Behaghel[a], Marc Ferrari[a], Wilfried Jahn[c], Christophe Gaschet[d], Bertrand Chambion[d], David Henry[d]

[a]Aix Marseille Univ, CNRS, CNES, LAM, Laboratoire d'Astrophysique de Marseille, 38, rue Joliot-Curie, Marseille,13388, France; [b]Kazan National Research Technical University named after A.N. Tupolev –KAI, 10 K. Marx, Kazan 420111, Russia; [c] Space Structure Lab, Caltech, 1200 E California Blvd, Pasadena, CA 91125, USA; [d]Univ. Grenoble Alpes, CEA, LETI, MINATEC campus, F38054 Grenoble, France

**ABSTRACT**

In the present paper we consider quantitative estimation of the tolerances widening in optical systems with curved detectors. The gain in image quality allows to loosen the margins for manufacturing and assembling errors. On another hand, the requirements for the detector shape and positioning become more tight. We demonstrate both of the effects on example of two optical designs. The first one is a rotationally-symmetrical lens with focal length of 25 mm, f-ratio of 3.5 and field of view equal to 72°, working in the visible domain. The second design is a three-mirror anastigmat telescope with focal length of 250 mm, f-ratio of 2.0 and field of view equal to 4°x4°. In both of the cases use of curved detectors allow to increase the image quality and substantially decrease the requirements for manufacturing precision

## 1. INTRODUCTION

The curved detectors technology is a rapidly developing area, which in the nearest future can make a huge impact on performance of imaging optical systems and approaches to designs them. A number of functional devices, produced by different techniques were demonstrated by a number of research groups over the world[1-5].

The key advantages of use of the curved detectors are were demonstrated in a number of researches[6-7]. Firstly, when there is no need to correct the field curvature, a better image quality can be achieved. Secondly, the required quality can be reached in a simpler optical system. Thirdly, due to change of the angle of incidence to the focal surface, the image illumination can be substantially increased.

Obviously, the use of a curved detector should also affect the tolerances analysis for an optical system. Better image quality and/or simplification of the optical system design in the nominal state implies wider margins for the manufacturing and assembling errors. On the other hand, the optical system should become more sensitive to the detector parameters like the radius of curvature and its' linear and angular position. Thus, the main goal of this study is to quantify the effect, which introduction of a curved detector makes on tolerance analysis, and exemplify it by certain optical designs.

To provide a visual demonstration of the changes in performance and tolerances we consider two completely different optical systems – an axisymmetric lens and an off-axis telescope. For each of the systems we provide the optical quality and tolerances estimations. Then we perform a re-optimization with a curved detector and repeat the analyses to show the obtained gain.

The paper is organized as follows: in Section 2 we describe the demonstrative optical designs, Section 3 provides the image quality estimations for each case, in Section 4 the tolerance analysis results are described and the conclusions on the study are given in section 5.

---

*eduard.muslimov@lam.fr; phone +33 4 91 05 69 18; lam.fr

## 2. OPTICAL DESIGNS DESCRIPTIONS

We consider two completely different optical systems in order to make our study more general. However, we make an accent on fast optics with relatively wide field of view, which definitely benefits more from use of a curved detector.

The first demonstrative example is a wide-field axisymmetric lens system; the second one is a fast off-axis unobscured telescope. The main optical parameters of the systems are given in Table 1 and a more detailed description is provided below.

Table 1. The main parameters of the demonstrative optical designs.

|  | Axisymmetric lens | Off-axis telescope |
| --- | --- | --- |
| Focal length, mm | 25 | 250 |
| F-number | 3.5 | 2 |
| Field of view, º | 72 | 4x4 |

### 2.1 Axisymmetric lens system

The lens under consideration is based on the design presented in apatent[8]. It is an all-spherical design consisting of 7 lens with 1 cemented pair and an internal aperture stop. As one can see, the lens operates with a wide field of view (FoV) and a moderate aperture. We consider 3 configurations of this design:

- Configuration A: the initial design with 7 lenses and a flat detector,
- Configuration B: the same design re-optimized with a curved detector. During the re-optimization all the radii and thicknesses were set variable, the focal length was limited, the total length and the illumination on the FoV edge were limited. Also necessary boundaries on the center and edge thicknesses were applied.
- Configuration C: a new design obtained from the Configuration C after removing of 2 lenses and iterative optimization. The optimization conditions were the same as in the case B, but the target illumination was increased.

General views of the resultant optical designs are shown on Fig.1.

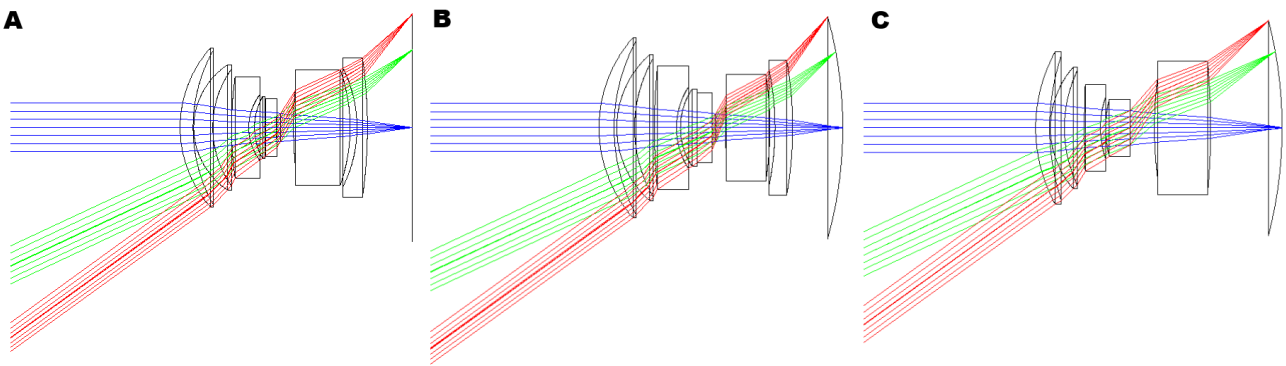

Figure 1. General view of the axisymmetric lens systems (f'=25 mm, F/3.5, 72º): A – initial configuration with 7 lenses and a flat detector; B – re-optimized configuration with 7 lenses and a curved detector; C – re-designed configuration with 5 lenses and a curved detector.

### 2.2 Off-axis telescope system

The off-axis telescope design represents a three-mirror anastigmate (TMA)[9]. It is an all-reflective system consisting of three aspherical mirrors with common axis and decentered apertures. The primary and secondary mirrors are conicoids, while the tertiary is described by the even asphere-type equation with 3 additional coefficients. Since there is no possibility exclude any optical surfaces or elements, we consider only 2 configuration, which definition is analogous to that given in p.2.1:

- Configuration A: the initial design with a flat detector,
- Configuration B: the same design re-optimized with a curved detector. In contrast with the previous example it is necessary to set limits on the marginal rays height in order to avoid obscuration or elements overlapping.

The general view of the optical system is shown on Fig.2.

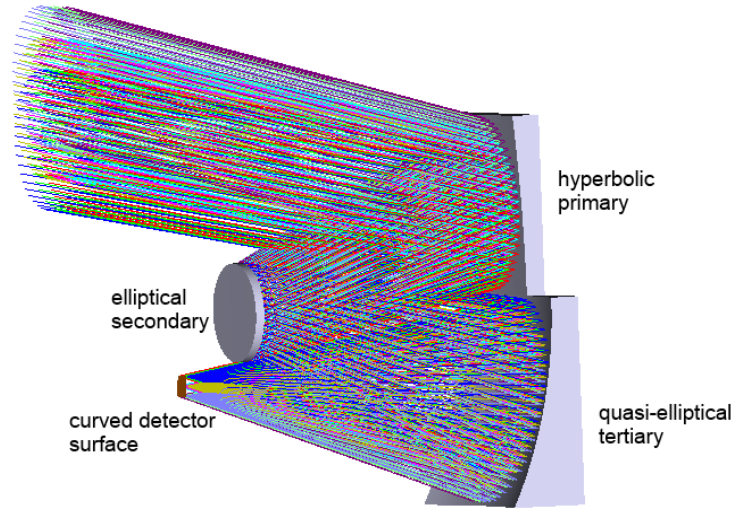

Figure 2. General view of the off-axis telescope (f'=250 mm, F/2, 4°x4°).

## 3. IMAGE QUALITY ANALYSIS

The image quality is assessed in a similar way for all the configurations of the both systems. We use spot diagrams and module transfer functions (MTF) to show the obtained in image quality.

### 3.1 Axisymmetric lens system

The spot diagrams for all the configurations of the demonstrative wide-field lens design are shown on Fig. 3. The root-mean square (RMS) radii of the spots are 9.6-33.6, 10.4-18.7 and 9.1-18.6 microns for the configurations A, B and C, respectively. One can see that use of the curved detector allows to obtain a notable increase of image quality at the FoV edge. Moreover, decrease of the optical components number almost doesn't change the geometrical aberrations.

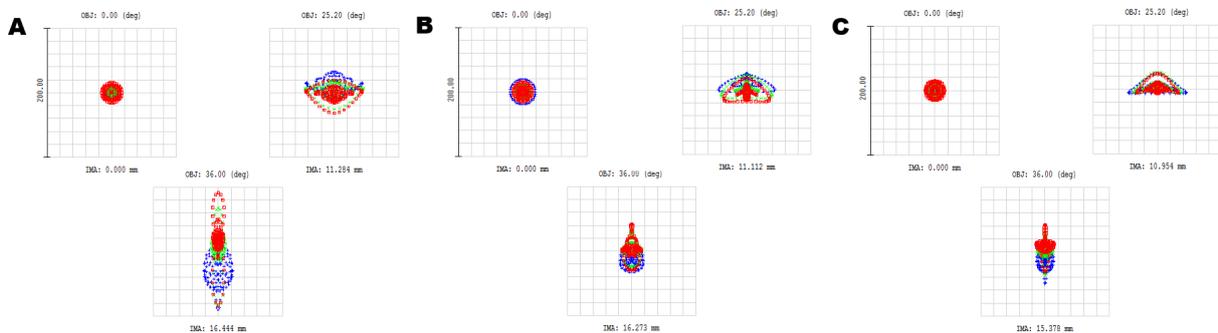

Figure 3. Spot diagrams (the colors correspond to wavelengths of 486, 587 and 656nm): A – initial configuration with 7 lenses and a flat detector; B – re-optimized configuration with 7 lenses and a curved detector; C – re-designed configuration with 5 lenses and a curved detector.

The MTF plots for the same designs are given on Fig. 4. They generally confirm the conclusions made on the basis of the spot diagram. We can even note that the configuration C has a higher resolution in the tangential direction in comparison with B.

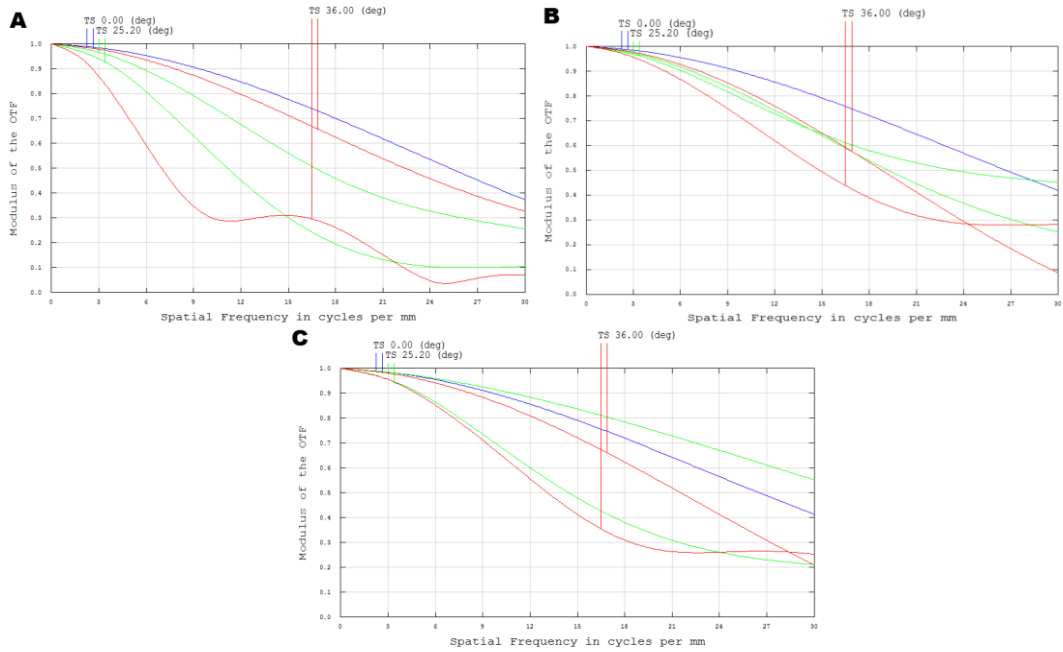

Figure 4. Module transfer functions (the colors correspond to field positions): A – initial configuration with 7 lenses and a flat detector; B – re-optimized configuration with 7 lenses and a curved detector; C – re-designed configuration with 5 lenses and a curved detector.

On top of the visual advantage in terms of the optical quality, the curved detector-based designs provide higher image illumination (see Fig. 5). Configuration B has a higher illumination value than the A configuration for the entire FoV except of its edge, while configuration C provides gain of a factor up to 1.7.

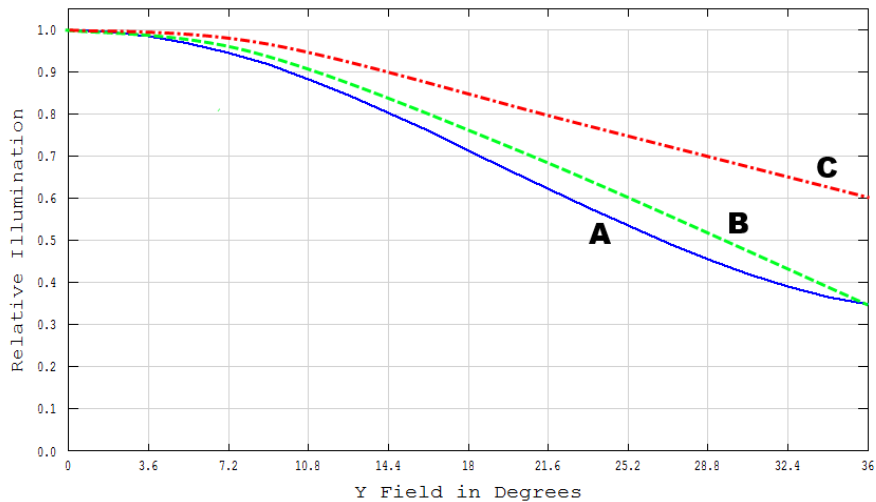

Figure 5. Relative illumination on the image surface: A – initial configuration with 7 lenses and a flat detector; B – re-optimized configuration with 7 lenses and a curved detector; C – re-designed configuration with 5 lenses and a curved detector.

## 3.2 Off-axis mirror system

Similarly, the telescopes spot diagrams are given on Fig.6. The RMS radii are 3.4-10.1 and 2.6-6.2 microns for the configurations A and B, respectively.

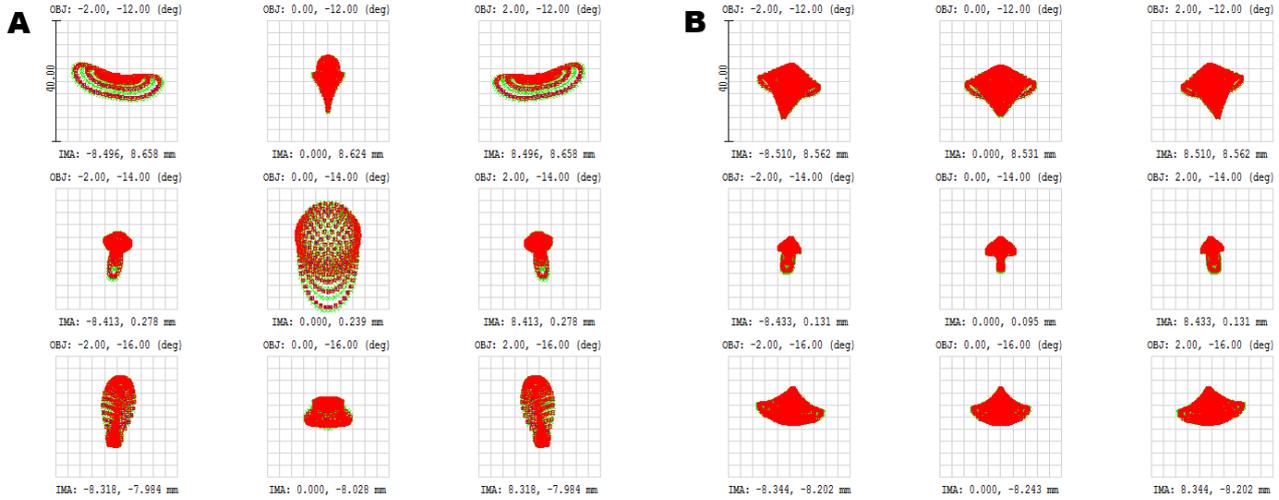

Figure 6. Spot diagrams: A – initial configuration with a flat detector; B – re-optimized configuration with a curved detector.

The MTF plots shows even larger difference in the image quality between the two designs (see Fig.7). Finally, we should mention that the difference in image illumination is negligible because the angular FoV is relatively small.

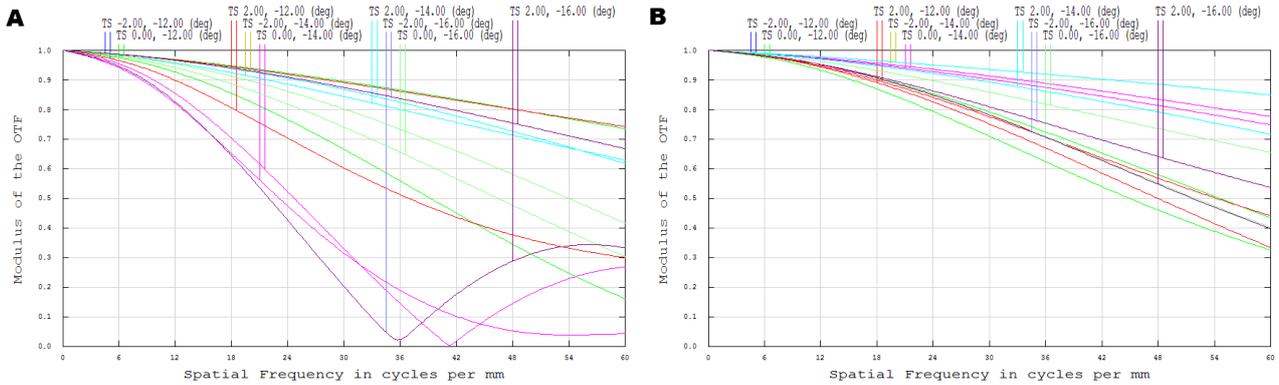

Figure 7. Module transfer functions (the colors correspond to field positions): A – initial configuration with a flat detector; B – re-optimized configuration with a curved detector.

## 4. TOLERANCE ANALYSIS

The tolerance analysis was performed in a similar way for all the designs. We make an assumption about the detector pixel size and require the spot RMS radii to be less than this value. We used only the standard analysis tools implemented in Zemax software. In order to account for simultaneous change of the design parameters we use the probability to achieve the target spot size found with the Monte-Carlo analysis as the main criterion. The aim is to guarantee the probability of 80% or more; if appears to be impossible with reasonable tolerances values, we tried to maximize the probability. For each of the systems the analysis was performed starting from the configuration providing the best quality. Then the tolerances were picked up to the next configuration and tightened or widened in an iterative process. In all the cases just a simple focus compensator is used and the reference wavelength is 532 nm. We must also note that truncation was applied when defining some of the tolerances.

### 4.1 Axisymmetric lens system

For the lens design we assume that the detector[10] has 24.6x24.6 mm$^2$ sensitive area with the pixel size of 24x24 μm$^2$. The demonstrative lens system contains 13 refractive surfaces, a physical diaphragm and the detector, so it is inconvenient to present the full tolerance results here. Instead we provide some analytics visualizing the obtained gain in the tolerances. Table 1 shows the maximum and minimum values of the tolerances describing the shape, refractive properties and positioning of all the optical surfaces. It's clear that the minimum values increase in the configurations with curved detectors. The gain can be as much as factor of 4 for the 7-lens design and factor of 40 for the 5-lens design.

Table 2. Extremal values of the tolerances for the axisymmetric lens.

|  | Configuration A | | Configuration B | | Configuration C | |
| --- | --- | --- | --- | --- | --- | --- |
|  | Min | Max | Min | Max | Min | Max |
| Radius, mm | 0.005 | 0.1 | 0.02 | 0.5 | 0.05 | 0.5 |
| Thickness, mm | 0.005 | 0.1 | 0.015 | 0.1 | 0.02 | 0.2 |
| Element decenter, mm | 0.02 | 0.1 | 0.02 | 0.1 | 0.05 | 0.1 |
| Element tilt, ° | 0.02 | 0.1 | 0.02 | 0.1 | 0.05 | 0.1 |
| Surface decenter, mm | 0.02 | 0.1 | 0.02 | 0.1 | 0.05 | 0.1 |
| Surface tilt, ° | 0.02 | 0.1 | 0.05 | 0.1 | 0.05 | 0.1 |
| Surface irregularity, μm | 0.03 | 0.25 | 0.06 | 0.25 | 0.02 | 0.22 |
| Refraction index | 0.0001 | 0.0001 | 0.0001 | 0.0001 | 0.004 | 0.004 |
| Abbe number | 0.05 | 1 | 0.1 | 2 | 1 | 2 |

The drawback of these change in tolerances is an expectable increase of requirements for the detectors shape and position. However, if we consider the quantitative change of these values (see Table 2) we find that: the tolerance on the curvature radius is relatively wide; the linear position should be more precise but the accuracy is achievable; the tolerances on tip and tilt becomes even wider because of the higher image quality. The only parameter which can become challenging is the local irregularity of the detector surface. The obtained values are definitely measurable, but their manufacturability strongly depends on the technology used.

Table 3. Detector parameters tolerances for the axisymmetric lens.

|  | Configuration A | Configuration B | Configuration C |
| --- | --- | --- | --- |
| Radius of curvature, mm | ∞ | 61.5 | 62.0 |
| Radius tol., mm | - | 1.0 | 1.0 |
| Irregularity tol., mm | 0.01 | 0.001 | 0.004 |
| Angular position tol., ° | 0.05 | 0.1 | 0.2 |
| Linear position tol., mm | 1 | 0.1 | 0.2 |

It should be emphasized that the obtained change in the tolerances is related not only to the smallest values, but for distribution of all the tolerances on design parameters. To visualize this change we use stacked histograms (see Fig.8-10). Note that the values are arranged into three groups, which can be referred to as "high precision", "moderate precision" and "standard precision". Also, some of the parameters are scaled to fit the same range. In general, one can note that with the increase of the image quality obtained due to introduction of the curved detectors all the tolerances gradually move from the "high/moderate precision" zone to the "standard precision" zone.

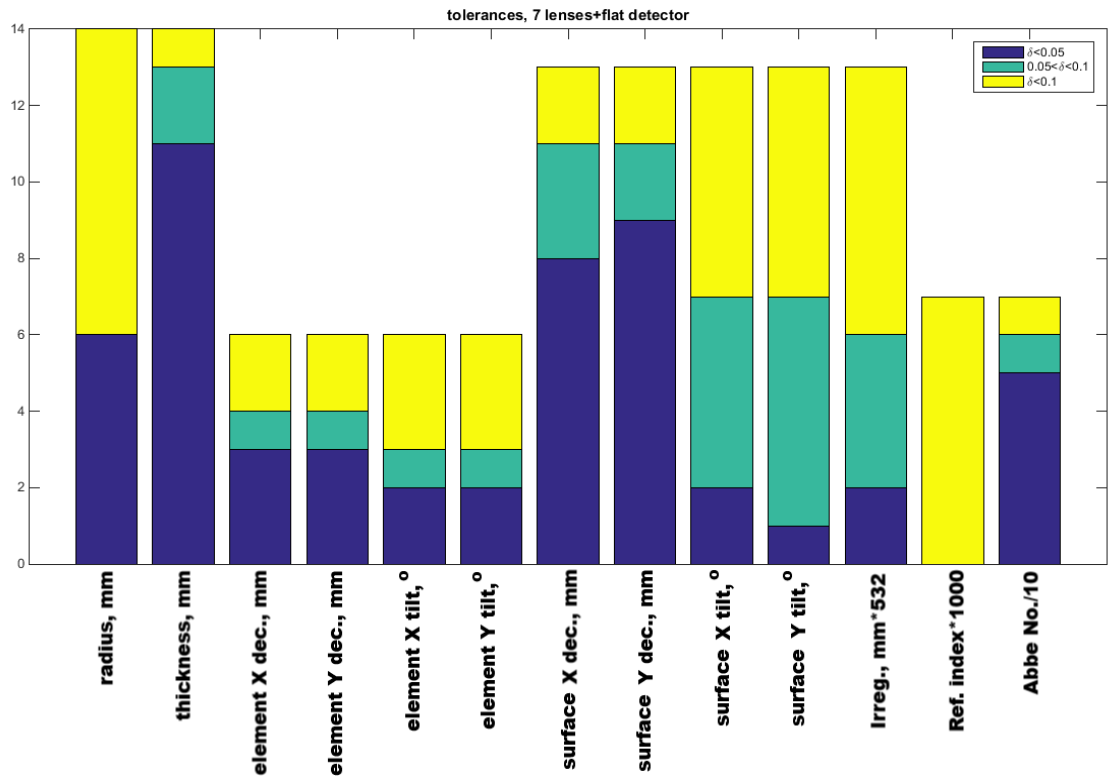

Figure 8. Tolerances overview for configuration A (initial configuration with 7 lenses and a flat detector).

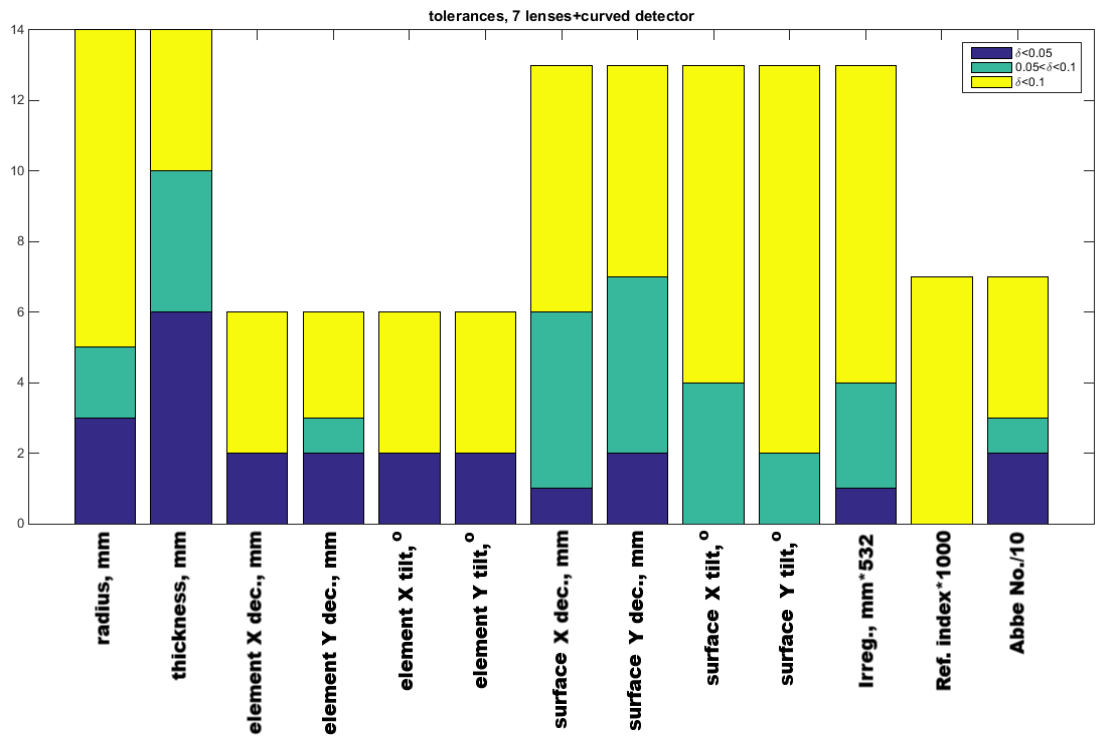

Figure 9. Tolerances overview for configuration B (re-optimized configuration with 7 lenses and a curved detector).

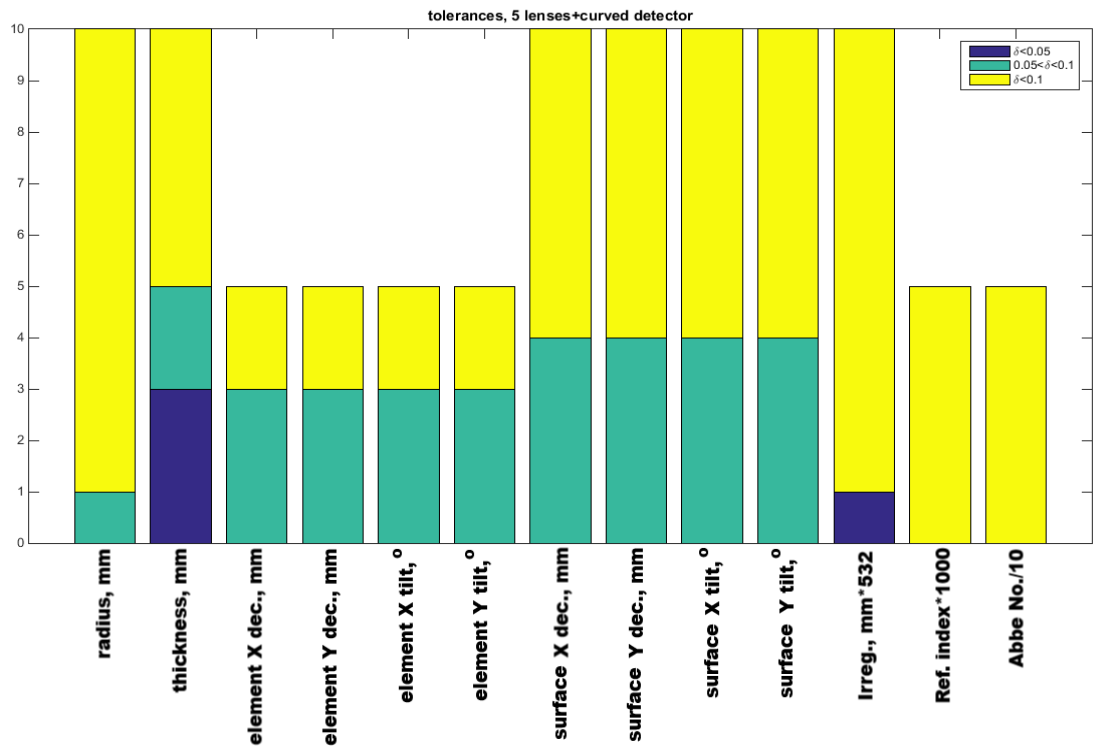

Figure 10. Tolerances overview for configuration C (re-designed configuration with 5 lenses and a curved detector).

Finally, Fig. 11 shows the Monte-Carlo analysis results. The grey line corresponds to the pixel size, the dotted vertical lines correspond to the nominal spot size increased by 5%. As one can see, the configuration A provides only 20% probability even with the strict manufacturing requirements. The configurations B and C are close to each other and completely fulfill our requirements to the probability.

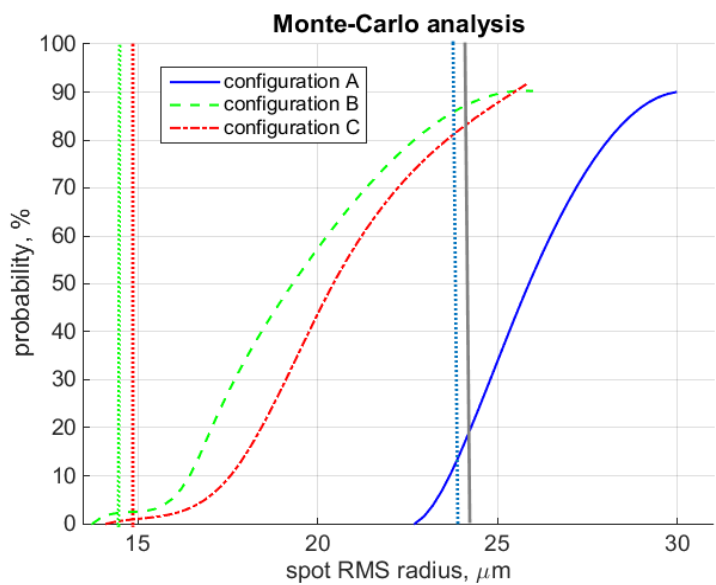

Figure 11. Results of the tolerances Monte-Carlo analysis for the axisymmetric lens

(the dotted lines mark the corresponding limits).

## 4.2 Off-axis mirror system

For the off-axis telescope we assume that the detector[11] has size of 17.7x17.7 mm² with a 11x11 µm² pixel. In this case we can provide the full description of the tolerancing results, since there are only three optical surfaces. They are presented in Table 4. As one can note, a trade-off of the tolerances takes place here. The largest values corresponding to the aspheres vertex radii becomes smaller, while the smallest one like the surface irregularity and angular misplacements become larger. The main point is that the change of the radii tolerance doesn't affect the manufacturing and testing complexity, while the change of the tilt and surface error margins make it much simpler.

Table 4. Values of the tolerances for the off-axis telescope.

|  | Radius, mm | Surface irreg., µm | Distance, mm | Decenter, mm | Tip/Tilt, ° |
|---|---|---|---|---|---|
| Configuration A ||||||
| Primary | 1.0 | 0.1 | 0.2 | 0.2 | 0.02 |
| Secondary | 1.0 | 0.007 | 0.2 | 0.2 | 0.05 |
| Tertiary | 0.6 | 0.1 | 0.2 | 0.1 | 0.01 |
| Configuration B ||||||
| Primary | 1.0 | 0.1 | 0.2 | 0.2 | 0.04 |
| Secondary | 0.5 | 0.05 | 0.2 | 0.2 | 0.05 |
| Tertiary | 0.2 | 0.1 | 0.2 | 0.1 | 0.03 |

In a similar way we provide the tolerances on the detector parameters (see Table 5). One can see that the values become slightly smaller, but nor of them approach the limits of manufacturability or testability.

Table 5. Detector parameters tolerances for the off-axis telescope.

|  | Configuration A | Configuration B |
|---|---|---|
| Radius of curvature, mm | ∞ | 709.4 |
| Radius tol., mm | - | 1.0 |
| Irregulerity tol., mm | 0.1 | 0.1 |
| Angular position tol., ° | 0.2 | 0.1 |
| Linear position tol., mm | 2 | 0.5 |

Finally, the Monte-Carlo results are given on Fig. 12. The conventions are the same as those used on Fig. 11. Both of the designs meet our requirements for the probability and both of the curves almost converge around the pixel size mark. However, the configuration B in general provides a higher probability to obtain the desired optical quality.

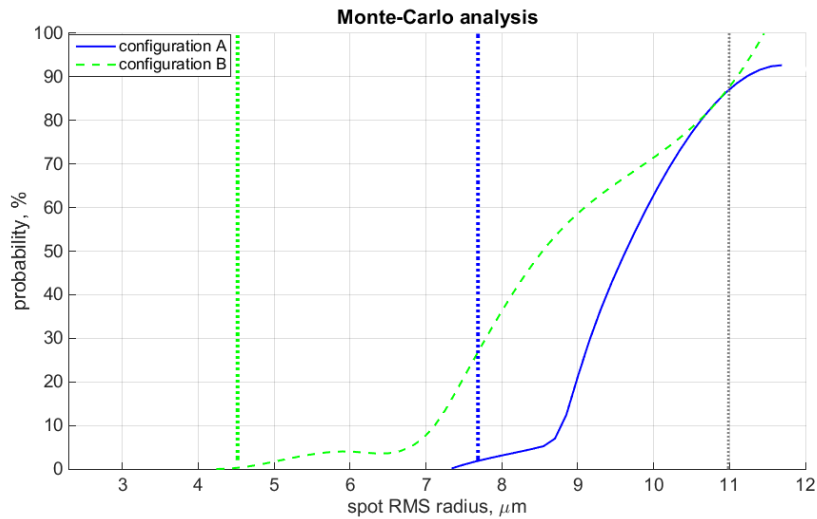

Figure 12. Results of the tolerances Monte-Carlo analysis for the off-axis telescope

(the dotted lines mark the corresponding limits).

## 5. CONCLUSIONS

In the present paper we explored the influence of introduction of curved detectors into wide-field imaging optical systems. The expected effects are substantial increase of the image quality which allows to widen the margins of the manufacturing precision requirements; and tightening of the tolerances on the shape and position of the detector itself. However, we clearly shown on examples of two different optical designs that the first effect is superior over the second one. While the requirements related to the detector remain feasible, the required level of precision for the optical elements can be decreased by an order of magnitude. Therefore, on top of the gain in the optical quality and image illumination the designs with curved focal planes have such an advantage as better manufacturability and simplicity of assembling. In our opinion, this last advantage is often missed in discussions of the curved detectors prospects. Particularly, this feature can change the estimative cost of building of a new optical system with a curved detector in comparison with a one using a standard flat detector. In general, we believe that development of the curved detectors technology and the corresponding optical design techniques can be useful for the future astronomical instruments like E-ELT HARMONI[12], MSE[13] or Hector[14].

One of the factors of risk, which was demonstrated in this study is the relatively high requirement to the detector's shape local errors occurring in some cases. The possibility to measure and control this value should be a subject of a separate research.

## ACKNOWLEDGEMENTS

The authors acknowledge the support of the European Research council through the H2020 - ERC-STG-2015 – 678777 ICARUS program. This research was partially supported by the HARMONI instrument consortium. We also would like to thank Pierre Baron from Altechna Company for fruitful discussions, which helped us to state the aim of this study.